\documentclass[showpacs, showkeys, amsmath, amssymb, notitlepage, superscriptaddress, reprint]{revtex4-2}

\usepackage[normalem]{ulem}
\usepackage[usenames,dvipsnames]{color}
\usepackage{upgreek}
\usepackage{dcolumn}

\usepackage{graphicx}
\usepackage{bm}
\usepackage{siunitx}
\usepackage[utf8]{inputenc}
\usepackage[T1]{fontenc}
\usepackage{mathptmx}
\usepackage{textcomp}
\usepackage{hyperref}

\begin{document}
	
	\title{Quantum emulation of the transient dynamics in the multistate Landau-Zener model}
	\author{Alexander Stehli}
	\affiliation{Institute of Physics, Karlsruhe Institute of Technology, 76131 Karlsruhe, Germany}
	\author{Jan David Brehm}
	\affiliation{Institute of Physics, Karlsruhe Institute of Technology, 76131 Karlsruhe, Germany}
	\author{Tim Wolz}
	\affiliation{Institute of Physics, Karlsruhe Institute of Technology, 76131 Karlsruhe, Germany}
	\author{Andre Schneider}
	\affiliation{Institute of Physics, Karlsruhe Institute of Technology, 76131 Karlsruhe, Germany}
	\author{Hannes Rotzinger}
	\affiliation{Institute of Physics, Karlsruhe Institute of Technology, 76131 Karlsruhe, Germany}
	\affiliation{Institute for Quantum Materials and Technologies, Karlsruhe Institute of Technology, 76344 Eggenstein-Leopoldshafen, Germany}
	\author{Martin Weides}
	\affiliation{James Watt School of Engineering, University of Glasgow, Glasgow G12 8LT, United Kingdom}
	\author{Alexey V. Ustinov}
	\affiliation{Institute of Physics, Karlsruhe Institute of Technology, 76131 Karlsruhe, Germany}
	\affiliation{Institute for Quantum Materials and Technologies, Karlsruhe Institute of Technology, 76344 Eggenstein-Leopoldshafen, Germany}
	
	\begin{abstract}
		Quantum simulation is one of the most promising near term applications of quantum computing. Especially, systems with a large Hilbert space are hard to solve for classical computers and thus ideal targets for a simulation with quantum hardware. In this work, we study experimentally the transient dynamics in the multistate Landau-Zener model as a function of the Landau-Zener velocity. The underlying Hamiltonian is emulated by superconducting quantum circuit, where a tunable transmon qubit is coupled to a bosonic mode ensemble comprising four lumped element microwave resonators. We investigate the model for different initial states: Due to our circuit design, we are not limited to merely exciting the qubit, but can also pump the harmonic modes via a dedicated drive line. Here, the nature of the transient dynamics depends on the average photon number in the excited resonator. The greater effective coupling strength between qubit and higher Fock states results in a quasi-adiabatic transition, where coherent quantum oscillations are suppressed without the introduction of additional loss channels. Our experiments pave the way for more complex simulations with qubits coupled to an engineered bosonic mode spectrum.
	\end{abstract}
	
	\maketitle
	\section*{Introduction}
	
	Noisy intermediate scale quantum (NISQ) devices, although imperfect, can have an advantage over classical computers at certain tasks \cite{Preskill2018, Arute2019}. Especially, analog quantum simulators (AQS), have advanced drastically over the last decade and can now tackle problems that are hard to solve for classical computers \cite{Bernien2017, Zhang2017a}. This is largely owed to the simplicity of their concept: in contrast to gate-based quantum computing, analog devices directly emulate a Hamiltonian of interest and thereby mimic all of its properties. Since they lack the flexibility and accuracy of a digital quantum computer, AQS are best suited to study universal effects, which are more or less resistant to small perturbations \cite{Preskill2018}. Therefore, open quantum systems are an appealing target for analog simulation, particularly, because they are also hard to model with both classical \cite{Mostame2017} and gate-based quantum computers \cite{Garcia-Perez2020}.
	
	One such system is the multistate Landau-Zener model. It describes the interaction of an excitation at a time-dependent energy level with at least one other state or mode spectrum. Examples are the underlying Hamiltonian models for molecular collisions \cite{Child1996} and chemical reaction dynamics \cite{Nitzan2006}. It is also ubiquitous in many artificial quantum devices, which have generated a broad research interest in recent years. In experimental studies on super- \cite{Oliver2005, Sillanpaa2006, Berns2008} and semiconducting qubits \cite{Petta2010, Ota2018}, or nitrogen-vacancy centers in diamond \cite{Childress2010, Fuchs2011} a Mach-Zehnder-like interference effect was exploited for quantum state preparations. The transient dynamics of the Landau-Zener model was also studied experimentally \cite{Yoakum1992, Berns2008, Zenesini2009, Huang2011}, albeit only for two-state systems. In contrast, various theoretical works on the topic focus on the more complex multistate Landau-Zener model \cite{Vitanov1999, Zueco2008, Orth2010, Orth2013}. In particular, the time evolution of the system's state and its dependence on the mode spectrum are of great interest there.
	
	In this work, we study the transient dynamics of the multistate Landau-Zener model of a superconducting quantum circuit. A tunable transmon qubit is used to realize a time-dependent energy state. An artificial spectral environment is formed by four harmonic oscillators which are implemented as lumped element microwave resonators. We study the time evolution of the system as a function of the Landau-Zener velocity $v$, i.e., the speed at which the energy level separation bypasses the narrow spectral bath. In circuit, we are not limited to exciting the system via the qubit, but can also pump the harmonic modes directly. This allows us to investigate the qubit's time evolution for various initial states, and, in particular, as a function of the average photon number in the system. In our experiments, we observe coherent oscillations of the qubit population. For low average photon numbers, we compare our findings with numerical QuTiP simulations. For a strongly driven bosonic bath the transient dynamics gradually changes from a coherent oscillating to an adiabatic transition of the mode ensemble. We identify the $\sqrt{n}$-enhancement of the coupling strength \cite{TC_model} between the qubit and resonators with the number of photons $n$ in the system as the underlying cause. 
	
	\section*{Results}
	\subsection*{Implementation of the simulator}
	The Hamiltonian of the multistate Landau-Zener model reads
	\begin{equation}
	\label{eq:LZ_H}
	\frac{\hat{H}_\text{LZ}}{\hbar} = \frac{v t}{2} \hat{\sigma}_z + \sum_{n} \left( \omega_n \hat{a}^\dagger_{n} \hat{a}_{n} + g_{n} \hat{\sigma}_x\left(\hat{a}^\dagger_{n} + \hat{a}_{n}\right) \right).
	\end{equation}
	An implementation of this system with superconducting devices is straightforward, due to their tailored functionality and the good control over all circuit parameters.
	\begin{figure*}
		\centering
		\includegraphics{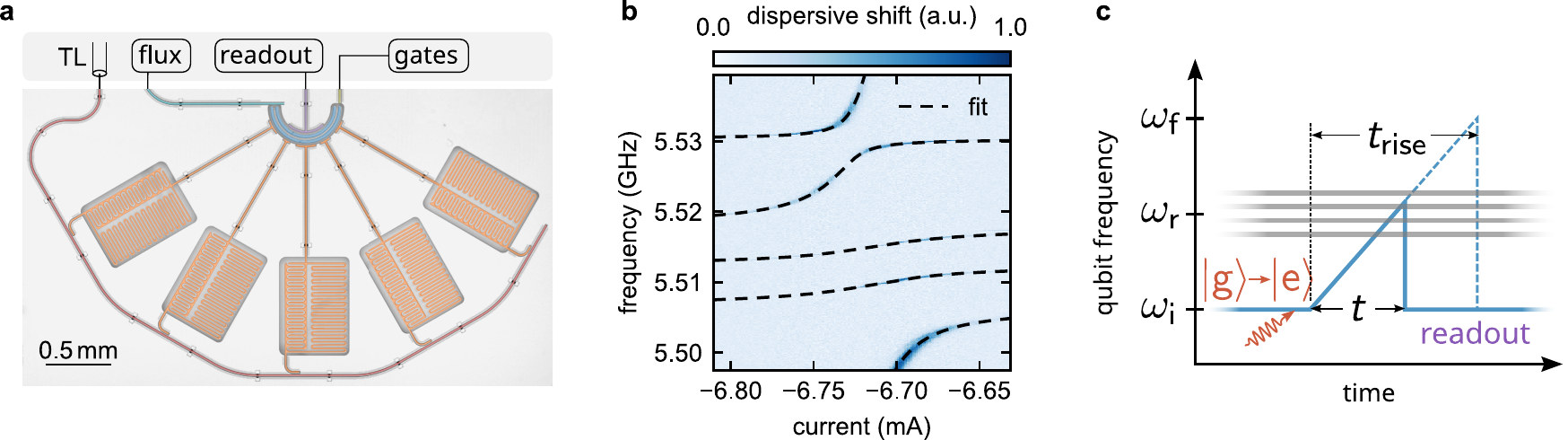}
		\caption{\textbf{Characterization of the quantum simulator and simulation scheme.} \textbf{a} Micrograph of the simulation device. An ensemble of five lumped element resonators (orange) with a dedicated input line (red) is coupled capacitively to a transmon qubit (blue). We achieve fast tunability using an impedance matched flux bias line (cyan) coupled to the SQUID of the qubit. DC and AC current biases, respectively generated by a current source and AWG, are combined with a bias tee. Qubit gates are admitted via a dedicated drive port (yellow). A $\lambda/4$-resonator (purple) is employed for the dispersive measurement of the qubit state. \textbf{b} Two-tone spectroscopy of the avoided level crossings between the first qubit transition and bosonic modes. The coupling coefficients are extracted from a fit to the mode spectrum, see black dashed lines. They are on the same order as the frequency spacing between the harmonic modes. \textbf{c} At the start of the quantum simulation, the qubit frequency $\omega_\text{q}$ (blue line) is tuned to $\omega_\text{i}$. Here, the systems initial state is prepared, shown for a $\pi$-pulse to the qubit. Thereafter follows a linear increase of the qubit frequency for a certain time $t$. We stop the qubit frequency sweep by returning the qubit to $\omega_\text{i}$, where dispersive state readout is performed. The qubit population along the trajectory is determined by repeating this process for different values of $t$ up to a maximum rise time $t_\text{rise}$, where $\omega_\text{q} = \omega_\text{f}$.}
		\label{fig:sample_char}
	\end{figure*}
	Figure~\ref{fig:sample_char}\textbf{a} displays a micrograph of our simulation device. We use a tunable transmon qubit to approximate the two level system with a time-dependent energy transition, i.e., the leading term in the Hamiltonian. An impedance matched flux bias line coupled to the SQUID of the qubit enables fast control of the qubit frequency. At the working point, the qubit is tunable over a frequency range of up to $\sim$\SI{400}{\mega\hertz}. This is partially limited by the bias tee employed to combine AC and DC current biases, see supplementary material. The qubit has an anharmonicity of $\alpha/2\pi = 241\,\si{\mega\hertz}$ and a rather large decoherence rate of $\Gamma_2 = 5\,\si{\micro\second}^{-1}$, which is not limited by the energy decay rate $\Gamma_1 = 0.2\,\si{\micro\second}^{-1}$. Due to the steep qubit dispersion at the operation point, flux noise is a likely limiting factor. The bosonic modes in the model are emulated by several lumped element microwave resonators with frequencies situated at around \SI{5.5}{\giga\hertz} and capacitively coupled to the qubit. Experimentally, we find that only four of the five resonators interact with the qubit, see supplementary material for detailed information. Qubit readout is performed dispersively utilizing a $\lambda/4$-stripline resonator with a resonance frequency of~$6.86\,\si{\giga\hertz}$. \cite{Blais2004, Wallraff2004}. Qubit gates are applied via a dedicated drive line. An additional input line coupled to the bath resonators allows for a direct spectroscopy of their transition frequencies and can be exploited in time domain experiments to directly apply microwave pulses to the bosonic modes.

	Figure~\ref{fig:sample_char}\textbf{b} shows a two-tone measurement of the coupled multi-mode system, in the single-photon manifold. Here, the qubit frequency is tuned in and out of resonance with the bosonic modes via a DC current bias applied to the flux coil. Thereby, several avoided level crossings are revealed, which we employ to extract the coupling coefficients between the qubit and resonators. The frequency spacing of the bosonic modes is of the same order as their coupling strength to the qubit. The extracted device parameters are summarized in the supplementary material and are also employed in the numerical simulation of the system.
	
	The Landau-Zener model is directly emulated by the quantum hardware. Therefore, no complex driving scheme is needed to run the simulation. The analog simulation follows the simple algorithm of system state preparation, free time evolution as the qubit frequency is increased, and finally qubit state readout, see Fig.~\ref{fig:sample_char}\textbf{c}. Due to experimental constraints, the qubit is initially flux biased to $\omega_\text{q} = \omega_\text{i}$ rather than to zero-frequency, approximately \SI{200}{\mega\hertz} below the bosonic modes' transitions. The Landau-Zener velocity is given by $v = (\omega_\text{f} - \omega_\text{i})/{t_\text{rise}}$, where the final qubit frequency $\omega_\text{f}$ is reached after the time $t_\text{rise}$. In the experiment, both $\omega_\text{i}$ and $\omega_\text{f}$ are measured quantities, while $t_\text{rise}$ is adjusted by the arbitrary waveform generator (AWG) used to tune the qubit frequency $\omega_\text{q}$. We determine the qubit population at several time steps along this trajectory. For state readout, the qubit is returned to $\omega_\text{i}$, halting the time evolution. The time resolution is limited to \SI{1}{\nano\second} by the AWG generating the tuning pulse.
	
	\subsection*{Multi-mode Landau-Zener model}
	
	For the first experiment discussed in this work, we prepare the system in the first excited state, i.e., the excited state of the qubit, by applying a weak $\pi$-pulse ($\sim$\SI{140}{\nano\second} length).
	\begin{figure*}
		\centering
		\includegraphics{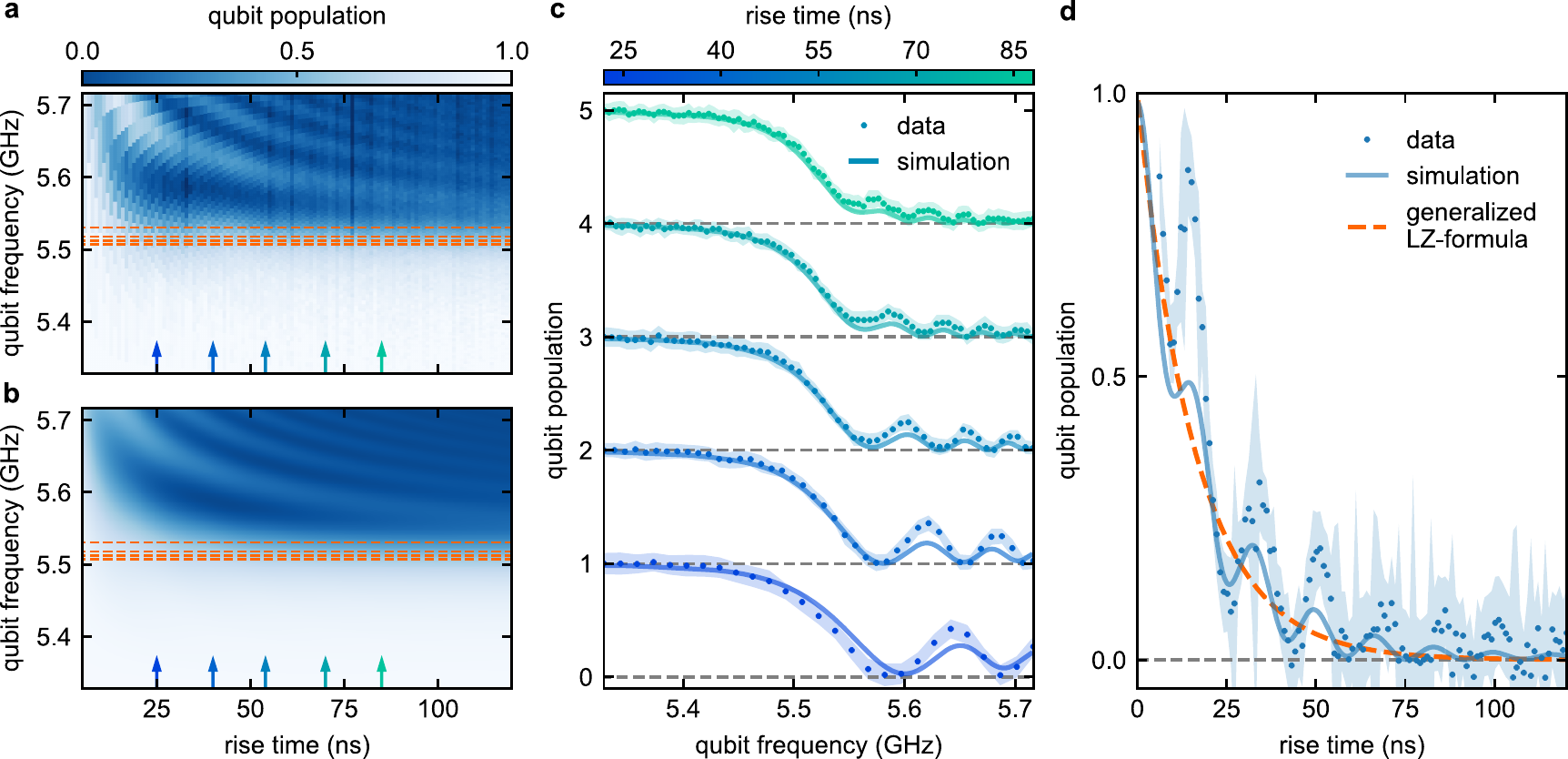}
		\caption{Transient dynamics in the multistate Landau-Zener model with an initially excited qubit. \textbf{a} Quantum and \textbf{b} numerical simulation of qubit population as a function of the qubit frequency $\omega_\text{q}(t) = \omega_\text{i} + vt$ and rise time $t_\text{rise} = (\omega_\text{q} - \omega_\text{f})/v$ with the Landau-Zener velocity $v$. During the transition of the resonator ensemble (frequencies indicated by dashed orange lines), the population of the initially excited qubit undergoes a steep drop, followed by coherent oscillations with diminishing amplitude. This is especially apparent for long rise times. \textbf{c} Quantitative comparison of the experiment (dots) and numerical simulation (lines) for different rise times, indicated by blue arrows in \textbf{b}. Shaded areas indicate the standard deviation of the qubit population. Neighboring traces are shifted by 1 to improve visibility. \textbf{d} Final qubit population as a function of the rise time $t_\text{rise}$. Oscillations persist in measurement and simulation, nevertheless, the final qubit population is well approximated by the generalized Landau-Zener formula, see Eq.~\ref{eq:LZ-formula}.}
		\label{fig:LZ_qubit}
	\end{figure*}
	Thereby, we make sure that the time development is dominated by the dynamics of the Landau-Zener model, rather than rotations of the qubit state stemming from the preparation in a system non-eigenstate. Figure~\ref{fig:LZ_qubit}\textbf{a} displays the results of our analog quantum simulation, side by side with numerical QuTiP simulations \cite{QuTiP2012, QuTiP2013}, see Fig.~\ref{fig:LZ_qubit}\textbf{b}. In the numerical simulations, we truncate the bosonic modes to the first two Fock states and omit energy relaxation and decoherence of the system.
	
	Qualitatively, we observe that, independent of the rise time, the qubit retains its excited state population until it reaches the resonator ensemble. After scattering on the bosonic modes, coherent oscillations of the qubit population emerge. The oscillation frequency is proportional to the energy difference of the qubit and bath states \cite{Oliver2005}. Therefore, it is independent of the coupling coefficients and solely depends on the velocity $v$ and increases with time. The oscillation amplitude decreases with time, i.e., for increasing qubit frequency. We emphasize that this effect is not introduced by the limited life-, and coherence time of the simulation device. Rather, it is how the system converges to a static state achieved in the limit of $t_\text{rise} \rightarrow \infty$. 
	We confirm the good quantitative agreement of numerics and experiment in a trace-wise comparison for several different rise times, see Fig.~\ref{fig:LZ_qubit}\textbf{c}. In the experiment, we have to account for a small offset of \SI{5.2}{\nano\second} to $t_\text{rise}$, see methods section, likely caused by a distortion of the tuning pulse after traversing the microwave lines to the sample.
	
	Figure~\ref{fig:LZ_qubit}\textbf{d} displays the final qubit population as a function of rise time, in good agreement with numerical simulations. A potential cause of the overshoot of the qubit population for short rise times is constructive Landau-Zener-St\"uckelberg interference, which can occur due to the finite time needed to return the qubit frequency to $\omega_\text{i}$. We also compare our results with the generalized Landau-Zener formula \cite{Shytov2004, Wubs2006, Saito2007}, which predicts a qubit population of
	\begin{equation}
	\label{eq:LZ-formula}
	P_\text{q} = \prod_n\exp\left({-\frac{2\pi}{v} g_n^2}\right).
	\end{equation}
	for $t_\text{rise} \rightarrow \infty$.	Each avoided level crossing can be interpreted as a beamsplitter for incoming photons \cite{Oliver2005}. Therefore, multiple crossings consecutively reduce the probability of the photon to remain in the qubit. Apart from the remnants of coherent oscillations in the measurement, our data are in agreement with the analytical formula. As predicted, the qubit remains in the excited state for $t_\text{rise}\rightarrow 0$, with the final state exponentially approaching the ground state in the adiabatic limit of $t_\text{rise}\rightarrow\infty$.
	
	\subsection*{Excited ensemble}
	\begin{figure*}
		\centering
		\includegraphics{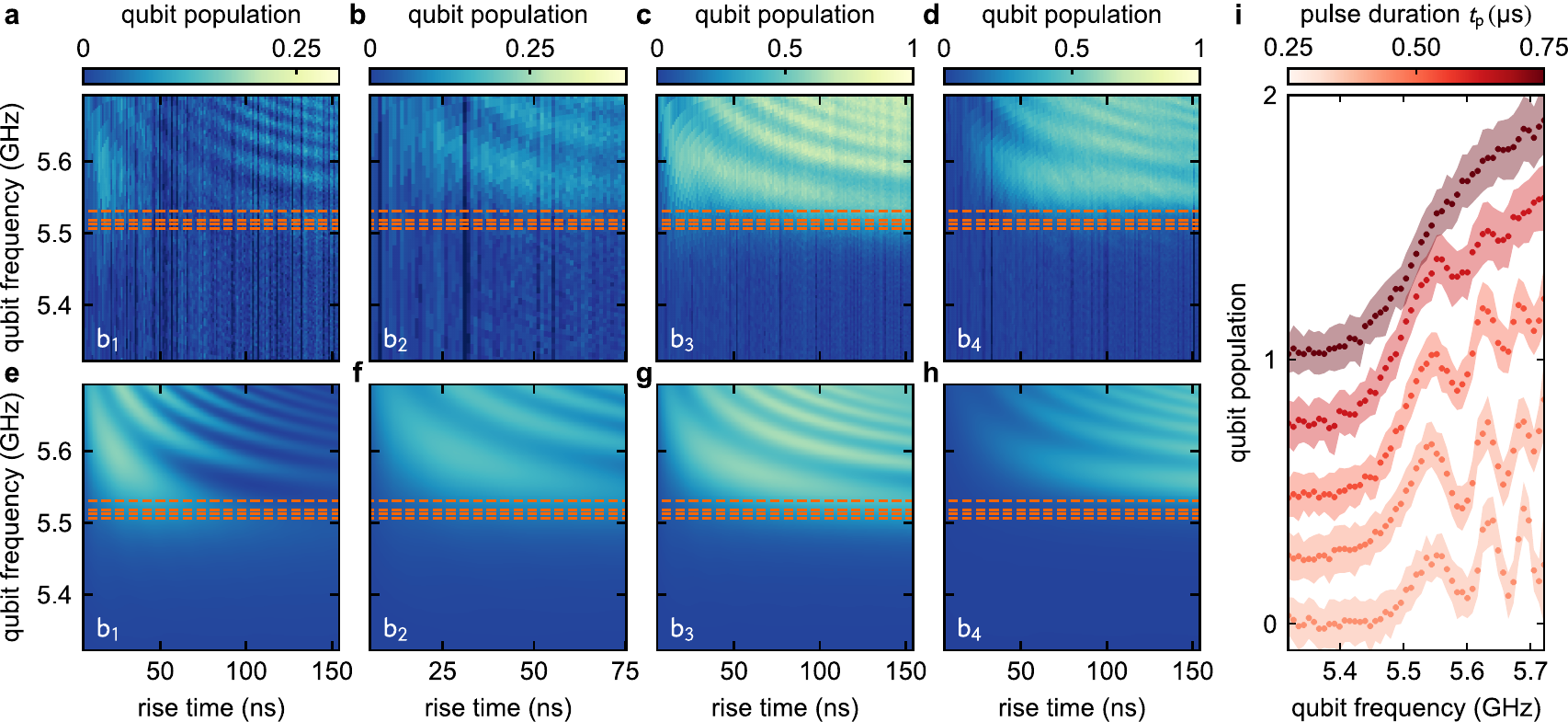}
		\caption{Qubit population dynamics with an excited resonator mode. The qubit starts in the ground state and transitions the bosonic modes, one of which is excited. Each row of the panels corresponds to a different initial state, where the resonators are excited in ascending order ($\text{b}_i$ for the $i$th resonator). In the experiment (\textbf{a}-\textbf{d}), a single resonator is prepared in a state with a low average population. (\textbf{e}-\textbf{f}) QuTiP simulation of the dynamics (compare \textbf{a} to \textbf{e}, etc.). We find coherent oscillations of the qubit population. As expected from a series of beamsplitters, the amplitude depends strongly on the excited resonator, i.e., where the photon is injected in the beamsplitter setup. \textbf{i} Transient dynamics for for $t_\text{rise} = 50\,\si{\nano\second}$ for a larger average photon number in resonator $\text{b}_1$. By increasing the duration $t_\text{p}$ (shown for 250, 300, 400, 550, and \SI{750}{\nano\second}) of the pump pulse higher Fock states contribute to the dynamics of the system. The $\sqrt{n}$-scaling of the effective coupling strength between qubit and resonator with the photon number $n$ of the respective Fock state increases the Landau-Zener tunneling probability, see Eq.~\eqref{eq:LZ-formula}. Initially, this results in a greater oscillation amplitude and a larger final value of the qubit population after the transit, but gradually shifts towards an adiabatic transition. For $t_\text{p} =750\,\si{\nano\second}$, the coherent oscillations of the qubit population are fully suppressed and the qubit is (approximately) in the excited state upon reaching $\omega_\text{f}$. Neighboring traces are shifted by 0.25 to improve visibility.}
		\label{fig:LZ_bath}
	\end{figure*} 
	Due to the dedicated drive line for the resonators, we can also pump the bosonic modes, rather than exciting the qubit. Thereby, their roles are reversed and the qubit, this time initially in the ground state, acquires (a fraction of) a photon from the populated mode during the transit. Since we cannot easily prepare the resonators in a Fock state, we apply a short drive pulse to one of the resonators. Thereby, it is prepared in a superposition state including higher levels, where the average photon number depends on the duration $t_\text{p}$ of the pump pulse. Since $t_\text{p}$ is lower than the resonators' decay rates $\kappa$ across all presented experiments, the pump does not result in a coherent state in the excited oscillator.
	
	Figure~\ref{fig:LZ_bath} displays the quantum (a-d) and numerical simulation (e-h) of the qubit population during the transition of resonator ensemble, one of which is excited. The pump pulse duration in the presented experiments was chosen as $t_\text{p} = 80 \,\si{\nano\second}$. In the numerical simulation, the bosonic mode was initialized in a one-photon Fock state to reduce the complexity of the calculation. Once again, coherent oscillations of the qubit population emerge after its frequency traverses the bath. Here, the amplitude of the oscillations strongly depends on the initial state of the system. This is particularly clear in the picture of a chain of beamsplitters, where the survival probability of an injected photon increases when injected at a later stage. Furthermore it also explains the shift of the "center of mass" of the oscillation towards larger $t_\text{rise}$ for a higher frequency of the initially excited resonator.
	
	What happens if we now increase the average photon number in the excited resonator? Figure~\ref{fig:LZ_bath}\textbf{i} shows the qubit population in dependence on the average photon number $\langle n \rangle$ in the lowest frequency resonator $\text{b}_1$ and for $t_\text{rise}=50\,\si{\nano\second}$. Here, we change $\langle n \rangle$ by varying the duration $t_\text{p}$ of the pump tone. Comparing with Fig.~\ref{fig:LZ_bath}\textbf{a}, a higher population of the resonator initially leads to more pronounced oscillations with a larger amplitude. For very long pump tones the behavior changes to an increasingly adiabatic transition of the avoided level crossing. In the Jaynes-Cummings model, which is the basis of our simulator, the matrix element between a Fock state with photon number $n$ and the qubit ground state scales with $\sqrt{n}$ \cite{TC_model}. The Landau-Zener tunneling probability scales with the square of this matrix element, see Eq.~\eqref{eq:LZ-formula}, which in turn corresponds to a faster (smaller Landau-Zener velocity $v$), more adiabatic transit. A direct comparison with numerical simulations proves challenging, due to the complex initial state of the pumped resonator and the large Hilbert space of the system. However, we can qualitatively reproduce the time evolution of the qubit population in a drastically simplified version of the system, comprising only two resonators, see supplementary material \ref{Appendix:large-photon-number}.
	
	\section*{Discussion}
	In this work, we experimentally investigated the time evolution of the multistate Landau-Zener model using a superconducting quantum simulator. The focus of this study was the transient dynamics of the qubit population in the vicinity of several bath resonators forming a spectral environment. By changing the Landau-Zener velocity $v$ and the initial state of the system we observed coherent oscillations of the qubit population, which subdue towards large $v$ where the transition becomes adiabatic. In contrast to previous studies of the model, our simulator allows for a preparation of higher excited states of the participating bosonic modes. For a larger average photon number $\langle n \rangle$, the transit of the avoided level crossings gradually becomes adiabatic, which is the result of an effectively larger coupling strength $g\propto\sqrt{n}$ between the qubit and the $n$'th Fock state of the excited resonator. Throughout the paper, we validate the results of the quantum experiments with numerical QuTiP simulations. Our implementation of an engineered spectral environment inspires future efforts towards emulating the dynamics of open quantum systems with superconducting quantum circuits. For example, it offers a promising approach to realize a quantum simulator of the spin boson model \cite{Leggett1987} with a tailored spectral function and, with the help of the additional drive tones, explore it in the ultra-strong coupling regime \cite{Ballester2012, Leppakangas2018}. 
	\section*{Methods}
	\subsection*{Dispersive measurement of the qubit population}
	
	As described in the main text, we determine the qubit state via the dispersive shift of the readout resonator. Since our measurement setup lacks a quantum limited amplifier, each data point is pre-averaged 500 times on the data acquisition card. This is repeated 25 times to calculate the standard deviation of the mean. Additionally, we had to calibrate the position of the resonator for ground and excited state. Especially, time dependent fluctuations of the resonator frequency would otherwise prove detrimental. Therefore, the measurement sequences for each time trace includes two additional points: For the first, we measure the dispersive shift after a $\pi$-pulse was applied to the qubit, which gives a reference point for the excited state. Second, the qubit frequency is tuned along the trajectory through the resonator ensemble and its state is measured after tuning it back to $\omega_\text{i}$. Since no excitation pulse is applied this gives a reference for the ground state. In the IQ-plane, we project all other data points in the sequence onto the connection line between the two reference measurements. The (effective) qubit population is then given by
	\begin{equation}
		\label{eq:calib}
		P_\text{q} = \frac{x_i - x_\text{g}}{x_\text{e} - x_\text{g}},
	\end{equation}
	where $x_{i}$ is the position of point $i$ along the line, and $x_\text{g/e}$ denotes the position of ground/excited state. The standard deviation is calculated from Gaussian error propagation. \newline
	\linebreak
	
	\subsection*{Offset calibration}
	We calibrate the offset of our time traces in the measurement by a comparison with the numerical simulations. Here, we minimize mean squared error between each data point and its corresponding point in the numerical simulation, as a function of a time and frequency offset to the qubit population. We find the lowest deviation for a shift of \SI{5.2}{\nano\second} and \SI{2.4}{\mega\hertz} of the quantum simulation with respect to the numerical results.
	\linebreak\linebreak
	
	\section*{Acknowledgments}
	Cleanroom facilities use was supported by the KIT Nanostructure Service Laboratory (NSL). We are grateful to Alexander Shnirman for fruitful discussions. This  work  was  supported  by  the  European Research  Council  (ERC)  under  the  Grant Agreement No.  648011, Deutsche Forschungsgemeinschaft (DFG) projects INST 121384/138-1FUGG and WE 4359-7, EPSRC Hub in Quantum Computing and Simulation EP/T001062/1, and the Initiative and Networking Fund of the Helmholtz Association. AS acknowledges support from the Landesgraduiertenf\"orderung Baden-W\"urttemberg (LGF), JDB acknowledges support from the Studienstiftung des Deutschen Volkes. TW acknowledges support from the Helmholtz International Research School for Teratronics (HIRST).
	\bibliography{LZ_bib}
	
	\clearpage

	\pagestyle{plain}
	\section*{Supplementary material}
	\renewcommand{\thepage}{S\arabic{page}}
	\renewcommand{\thesection}{S\arabic{section}}
	\renewcommand{\thetable}{S\arabic{table}}
	\renewcommand{\thefigure}{S\arabic{figure}}
	\setcounter{figure}{0}
	\setcounter{page}{1}
	\setcounter{section}{0}
	
	\renewcommand\thesection{\Alph{section}}
	
	\section{Characteristics of the resonator ensemble}
\begin{figure}[h]
		\includegraphics{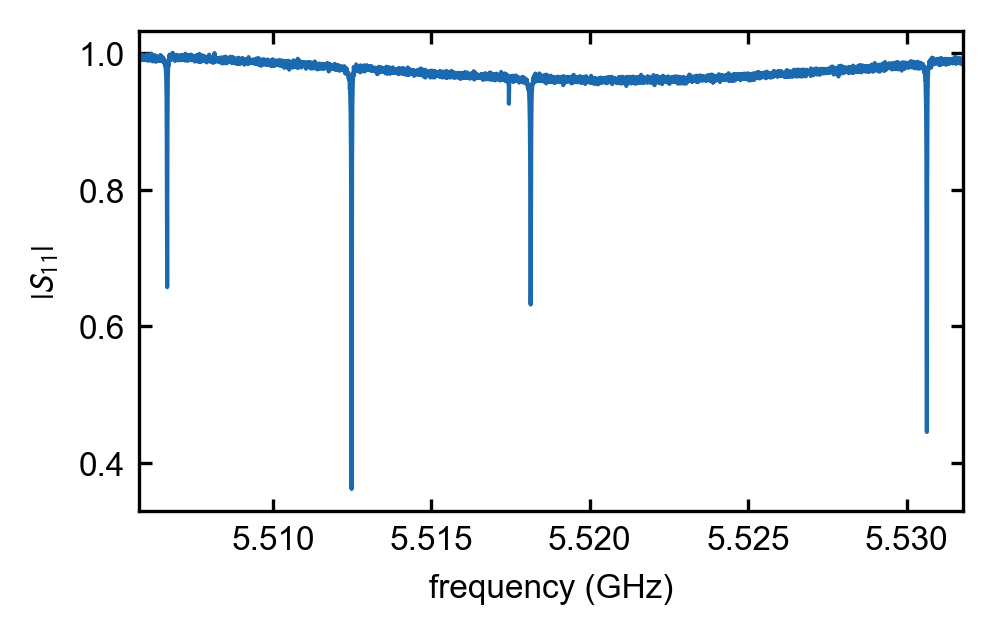}
		\caption{.}
		\label{fig:bath_scan}
	\end{figure}

	Figure~\ref*{fig:bath_scan} displays the reflection coefficient $\left\lvert S_{11}\right\rvert$ of the resonator ensemble. Here, the qubit is far detuned (on the scale of the coupling strength) from the bath resonators. There are five absorption dips, one for each physical resonator. The resonance in the center is barley coupled to the transmission line and does not couple to the qubit. This indicates that two of the physical resonators have hybridized, forming an asymmetric and a symmetric mode. Due to symmetry, only the latter interacts with the qubit, however, with a greater coupling strength than that of the individual modes.
	
	Using a circle fit \cite{Probst2015} to the complex valued $S_{11}$, we calculate the loaded and internal quality factor, $Q_\text{L}$ and $Q_\text{i}$, see table~\ref{table:bath_pars}.
	We extract the coupling strength between qubit an the resonator ensemble from a two tone measurement, see Fig.~\ref{fig:sample_char}. By applying a current to the local flux bias line, the qubit is tuned to resonance with the ensemble. The transverse capacitive coupling leads to a collective avoided level crossing of all participating modes. We determine the coupling coefficients in a fit, neglecting residual resonator-resonator couplings, and approximating the qubit dispersion as linear in the vicinity of the crossing. All system parameters are summarized in table~\ref{table:bath_pars}.
	
	\begin{table}[h]
	\caption{Properties of the resonator ensemble.}
	\begin{tabular}{c c c c r}
		resonator & $\omega/2\pi$ (GHz)& $Q_\textbf{i}$ $\left(\times 10^3\right)$ & $Q_\textbf{L}$ $\left(\times 10^3\right)$ & g/$2\pi$ (MHz)\\ \hline
		b$_1$& 5.507 & 402 & 329 & 14.6 \\
		b$_2$& 5.513 & 368 & 251 & 12.1 \\
		b$_3$& 5.518 & 268 & 221 & 14.4 \\
		b$_4$& 5.531 & 396 & 287 & 6.3 \\
	\end{tabular}
	\label{table:bath_pars}
	\end{table}
	
	\section{Bias tee calibration}
\begin{figure}
		\centering
		\includegraphics{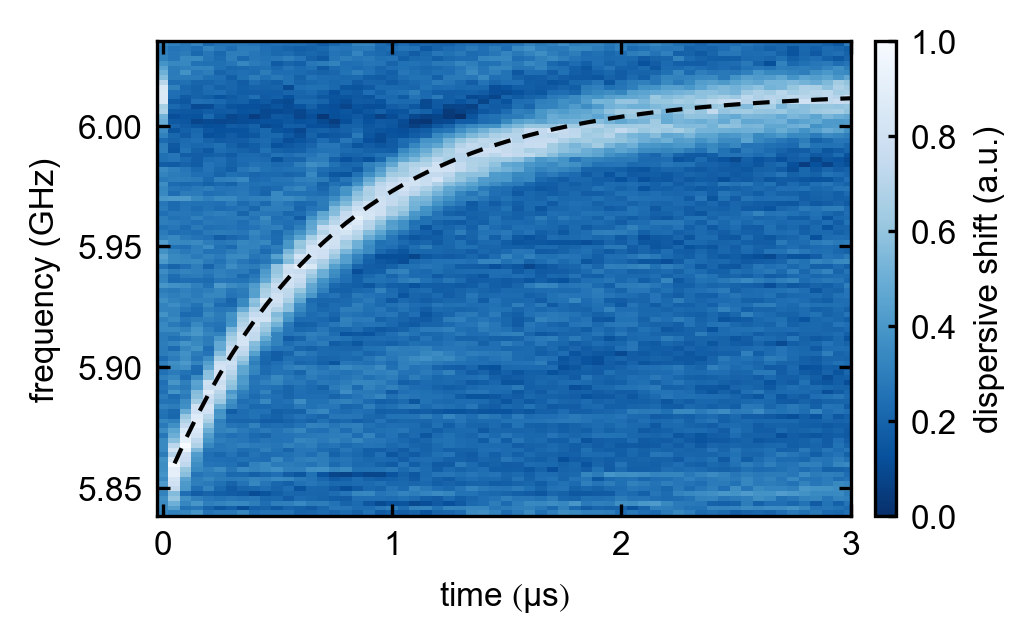}
		\caption{Bias tee calibration. $\pi$-pulse spectroscopy with an additional rectangular current pulse. The qubit frequency follows an exponential curve (black dashed line), which directly yields the time constant of the bias tee.}
		\label{fig:BiasT_calib}
	\end{figure}

	We employ a bias tee to combine ac and dc current biases. Due to the finite capacitance/inline resistance, current pulses applied to the bias tee decay exponentially with time. The time constant can be extracted by applying a rectangular current pulse to the bias tee, during which a $\pi$-pulse spectroscopy is performed on the qubit, see Fig.~\ref{fig:BiasT_calib}. Increasing the time delay between $\pi$-pulse and current pulse reveals the trajectory of the qubit during the current pulse. Since the qubit dispersion is approximately linear at the bias point, the qubit frequency follows an exponential curve with the time constant of the bias tee. For our setup, we find a time constant of $\tau = (718 \pm 6)\,\si{\nano\second}$. In the experiments, this is accounted for by adjusting the voltage pulse generated by the AWG to
	\begin{equation}
		\label{eq:biastee_V}
		V_\text{AWG} = R \left(I_\text{q} + \frac{1}{\tau} \int I_\text{q} \text{d}t\right),
	\end{equation}
	where $I_\text{q}$ is the designated current pulse at the qubit and $R$ is the inline resistance of the setup.
	
	\section{Numerical simulations with large photon numbers}
	\label{Appendix:large-photon-number}
\begin{figure}
		\centering
		\includegraphics{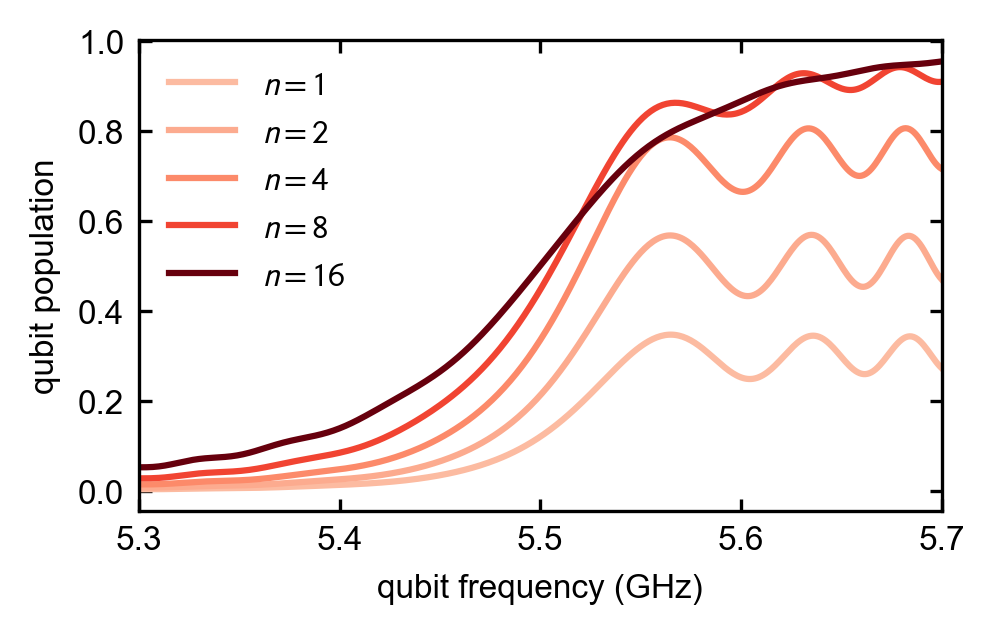}
		\caption{Numerical simulation of the Landau-Zener model with higher Fock states. The calculations comprise two resonators, where the lower frequency one is prepared in the $n$'th-Fock state.}
		\label{fig:Fockpump}
	\end{figure}

	Due to the significant size of the Hilbert space, numerical simulations with large average photon numbers in at least one of the resonators are not practicable. In order to validate the results of our experiment displayed in Fig.~\ref{fig:LZ_bath}\textbf{i}, we limit the simulation to two resonators with 20 Fock states each. Otherwise, the oscillators retain the properties of $b_1$ and $b_2$, see table~\ref{table:bath_pars}. The results of the numerics displayed in Fig.~\ref{fig:Fockpump} are in a good qualitative agreement with our quantum simulation.
	
	\clearpage
	
\end{document}